\documentclass[letterpaper]{article} 
\usepackage{aaai25}  
\usepackage{times}  
\usepackage{helvet}  
\usepackage{courier}  
\usepackage[hyphens]{url}  
\usepackage{graphicx} 
\usepackage{natbib}  
\usepackage{caption} 
\usepackage{algorithm}

\usepackage{newfloat}
\usepackage{listings}
\DeclareCaptionStyle{ruled}{labelfont=normalfont,labelsep=colon,strut=off} 
\floatstyle{ruled}
\newfloat{listing}{tb}{lst}{}
\floatname{listing}{Listing}
\usepackage{soul}
\usepackage{url}
\usepackage[utf8]{inputenc}
\usepackage{graphicx}
\usepackage{amsmath}
\usepackage{amsthm}
\usepackage{booktabs}

\usepackage[switch]{lineno}

\usepackage{pifont}
\usepackage{bbding}
\usepackage{fontawesome}

\usepackage{mdframed}

\usepackage{enumitem} 

\usepackage{algpseudocode}

\newcommand{\name}{{\sc {SpeHeatal}}}
\newcommand{\cname}{{\sc {Con2Dis}}}

\title{{\name}: A Cluster-Enhanced Segmentation Method for \\ Sperm Morphology Analysis}
\author{
    Yi Shi\textsuperscript{\rm 1,2}\equalcontrib \thanks{Corresponding author.},
    Yun-Kai Wang\textsuperscript{\rm 1,2}\equalcontrib,
    Xu-Peng Tian\textsuperscript{\rm 1,2},
    Tie-Yi Zhang\textsuperscript{\rm 1,2},
    Bing Yao\textsuperscript{\rm 3,4},
    Hui Wang\textsuperscript{\rm 3,4},\\
    Yong Shao\textsuperscript{\rm 3,4},
    Cen-Cen Wang\textsuperscript{\rm 3,4},
    Rong Zeng\textsuperscript{\rm 3,4}
}
\affiliations{
    \textsuperscript{\rm 1} School of Artificial Intelligence, Nanjing University, Nanjing, China\\
    \textsuperscript{\rm 2} State Key Laboratory for Novel Software Technology, Nanjing University, Nanjing, China\\
    \textsuperscript{\rm 3} Jinling Hospital, Affiliated Hospital of Medical School, Nanjing University, Nanjing, China\\
    \textsuperscript{\rm 4} Jiangsu Provincial Medical Key Discipline Cultivation Unit, Nanjing, China\\
    shiy@lamda.nju.edu.cn
}

\usepackage{bibentry}

\begin{document}

\maketitle

\begin{abstract}
The accurate assessment of sperm morphology is crucial in andrological diagnostics, where the segmentation of sperm images presents significant challenges. Existing approaches frequently rely on large annotated datasets and often struggle with the segmentation of overlapping sperm and the presence of dye impurities. To address these challenges, this paper first analyzes the issue of overlapping sperm tails from a geometric perspective and introduces a novel clustering algorithm, {\cname}, which effectively segments overlapping tails by considering three essential factors: \textbf{Con}nectivity, \textbf{Con}formity, and \textbf{Dis}tance. Building on this foundation, we propose an unsupervised method, {\name}, designed for the comprehensive segmentation of the \textbf{SPE}rm \textbf{HEA}d and \textbf{TA}i\textbf{L}. {\name} employs the Segment Anything Model~(SAM) to generate masks for sperm heads while filtering out dye impurities, utilizes {\cname} to segment tails, and then applies a tailored mask splicing technique to produce complete sperm masks. Experimental results underscore the superior performance of {\name}, particularly in handling images with overlapping sperm.

\end{abstract}

\section{Introduction}\label{sec:intro}

\begin{figure}[t]
	\centering
	\begin{minipage}[c]{\linewidth}
	\includegraphics[width=\linewidth]{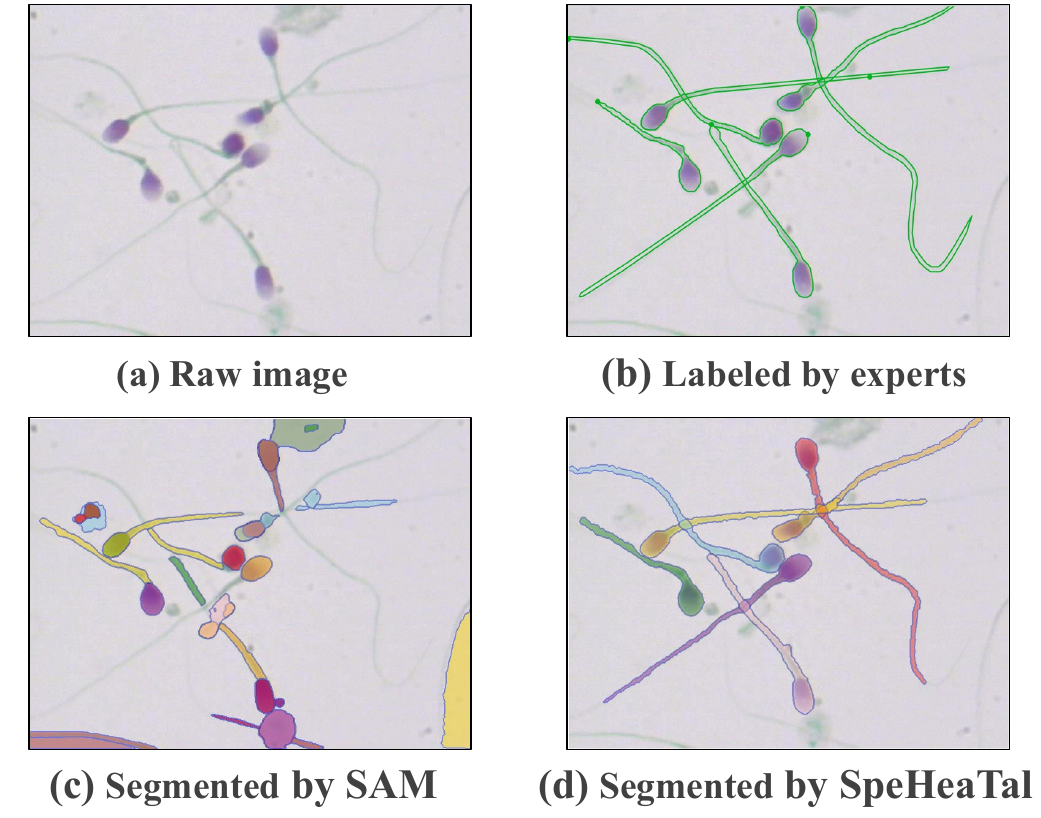}\\
    \end{minipage}
        \vspace{-4mm}
	\caption{
 Significant sperm overlap and dye impurities are evident in clinical practice. Compared with previous methods, such as SAM, our proposed {\name} effectively addresses these two challenges, achieving complete segmentation of multiple independent sperm.
  }\label{fig:intro}
        \vspace{-2mm}
\end{figure}

The analysis of sperm morphology constitutes a critical component within the framework of andrological diagnostics, serving as a key indicator of sperm viability and male reproductive capacity~\cite{Moretti2022Relevance,menkveld2011measurement,oehninger2021sperm,gatimel2017sperm}. Traditionally, the evaluation of sperm morphology has relied heavily on the expertise of andrologists, who conduct detailed examinations of semen samples under microscopy to distinguish between normal and abnormal sperm structures. This conventional approach not only imposes significant time demands on medical professionals but also fails to meet the growing need for high-throughput, efficient analysis. Compounded by the inherent subjectivity and the absence of uniform evaluative benchmarks, sperm morphology assessment emerges as a particularly challenging domain within reproductive diagnostics. Consequently, there is an increasing demand for automated detection systems capable of identifying sperm abnormalities. A primary challenge in this context is the accurate segmentation of individual spermatozoa within imaging, a task that is complicated by frequent sperm overlap, low contrast, and the presence of dye impurities in microscopic imagery. These factors exacerbate the segmentation challenge for current imaging models and complicate manual annotation, resulting in a marked scarcity of labeled sperm image datasets.

Recent advancements in image segmentation technologies have been widely applied across various domains\cite{DBLP:conf/aaai/FangZCSWL24,DBLP:conf/aaai/HeCFFMS24}, including medical imaging\cite{DBLP:conf/aaai/BiYZJHL024,DBLP:conf/aaai/Dai0LSW024,ma2024segment}. In the specific context of sperm morphology segmentation, existing models generally require labeled data for training. However, due to the limited availability of labeled sperm image datasets, these models often simplify the objects to be segmented, such as by only segmenting sperm heads or ignoring overlapping sperm. The ability to accurately segment {\em overlapping} spermatozoa {\em without} relying on labeled data is recognized as a crucial step in applying artificial intelligence to sperm morphology assessment.

Sperm image segmentation, as a task of instance segmentation, requires the segmentation of individual sperm masks without the necessity of identifying semantic differences among them. While recent advancements have shown significant progress in unsupervised instance segmentation~\cite{DBLP:journals/corr/abs-2312-17243,DBLP:conf/cvpr/0007GYM23,Alexander2023SAM}, their effectiveness is primarily confined to common instances in everyday scenes, with limited applicability to specialized targets such as sperm. For instance, our experimental results indicate that SAM~\cite{Alexander2023SAM} struggles to accurately identify the elongated structure of sperm tails, particularly in cases involving overlap and dye impurities, as depicted in Figure~\ref{fig:intro}(c). Additionally, some approaches have explored the use of clustering methods, such as K-means~\cite{macqueen1967some} and Spectual Clutering~(SC)~\cite{shi2000normalized}, for unsupervised instance segmentation, yet these methods also fail to adequately address the issue of overlap. While tubular structure segmentation~\cite{DBLP:conf/nips/Hu22,DBLP:conf/iccv/QiHQZY23} can deal with sperm tails, these methods treat overlapping sperm as a single entity rather than segmenting each sperm individually.

In response to these challenges, we first conduct a geometric analysis of sperm tail overlap, concluding that effective distinction of tails requires consideration of three key factors: \textbf{Con}nectivity, \textbf{Con}formity, and \textbf{Dis}tance. Based on these insights, we propose a novel clustering method, {\cname}, designed to achieve effective segmentation of sperm tails, with particular efficacy in overlapping regions. To the best of our knowledge, {\cname} is the first clustering algorithm explicitly developed to address the segmentation challenges posed by intersecting slender structures in sperm imagery. This approach may also offer valuable insights for the segmentation of images with similar structural characteristics, such as blood vessels and neural pathways.

Building on the foundation of {\cname}, we present {\name}, an end-to-end segmentation algorithm designed to accurately segment complete sperm. {\name} functions through a ``decomposition-combination'' approach, dividing the sperm segmentation task into distinct head and tail segmentation processes, which are subsequently integrated. Specifically, {\name} employs SAM to generate masks for sperm heads while filtering out dye impurities, leverages {\cname} to segment tails, and finally applies a tailored mask splicing technique to produce complete sperm masks. As an unsupervised algorithm, {\name} effectively mitigates the challenges posed by sperm overlap and dye impurities, delivering multiple independent and complete sperm masks without the need for any annotations.

Given that existing sperm datasets do not adequately capture the characteristics of sperm overlap and dye impurities observed in clinical practice, we collaborated with several prominent hospitals to compile a dataset comprising approximately 2,000 unlabeled sperm images for model calibration, along with an additional 240 expert-annotated images for rigorous model evaluation, as depicted in Figure~\ref{fig:intro}(b). Experimental results demonstrate the superior performance of {\name}, particularly in the context of overlapping sperm, as illustrated in Figure~\ref{fig:intro}(d). This advancement not only lays a foundation for future work in sperm morphology classification but also signals a significant step towards fully automated sperm morphology analysis using AI methodologies. Our contributions are summarized as follows:
\begin{itemize}
\item We introduce a novel clustering method, {\cname}, which effectively segments overlapping sperm tails by considering Connectivity, Conformity, and Distance.
\item We propose {\name}, a method capable of achieving effective and complete segmentation of overlapping sperm images under unsupervised conditions.
\item Experimental results on our clinical sperm dataset highlight the effectiveness of {\name}, especially in images with overlapping structures.
\end{itemize}

\section{Related Work}\label{sec:related}

\noindent\textbf{Sperm image segmentation} is a pivotal task in automating sperm morphology analysis. Existing studies, such as~\cite{DBLP:journals/npl/LvYQLZZ22,Iqbal2020Deep}, primarily focus on the segmentation of sperm heads, often neglecting the tails. In contrast, research efforts in~\cite{DBLP:journals/mbec/IlhanSSA20,DBLP:conf/hsi/FraczekKMLJM22} have concentrated on segmenting independent, non-overlapping sperm. Other approaches, such as those in~\cite{DBLP:journals/iet-cvi/KheirkhahMS19,DBLP:journals/cbm/LewandowskaWMLJ23}, treat overlapping sperm as a single entity, thereby neglecting the need to segment individual spermatozoa. These methodologies often simplify the segmentation task, leading to suboptimal results when applied to real-world clinical specimens. Moreover, these approaches typically require labeled datasets for training, which are limited in availability. \\

\noindent\textbf{Unsupervised instance segmentation} has seen significant advancements in recent years. The Segment Anything Model~(SAM)~\cite{Alexander2023SAM} demonstrates high-quality segmentation across a wide range of objects using diverse prompts without requiring specific fine-tuning. U2Seg~\cite{DBLP:journals/corr/abs-2312-17243} employs a hierarchical feature extraction framework to capture multi-scale information, thereby improving segmentation accuracy. Similarly, CutLER~\cite{DBLP:conf/cvpr/0007GYM23} utilizes contrastive learning and local feature representations to distinguish individual instances. However, these methods encounter challenges when applied to medical imaging due to variations in color spectrum, texture, and contrast specific to this domain. Their performance is particularly inadequate when addressing sperm overlap and dye impurities. \\

\noindent\textbf{Clustering methods for segmentation} have been widely integrated with advanced techniques to enhance image segmentation tasks. K-Means~\cite{macqueen1967some} optimizes segmentation by iteratively minimizing within-cluster variance, primarily focusing on the distance between pixels. 
Agglomerative Hierarchical Clustering~(AHC)~\cite{sneath1973numerical} constructs a hierarchy by iteratively merging the closest pairs of clusters from bottom to up, primarily focusing on distance and connectivity. Some clustering methods incorporate both distance and conformity, such as Local Subspace Affinity~(LSA)~\cite{DBLP:conf/eccv/YanP06} and Spectral Clustering~(SC)~\cite{shi2000normalized}. LSA computes an affinity between pairs of points based on principal subspace angles across different linear manifolds, while SC leverages the eigenvalues of a similarity matrix to reduce dimensionality before clustering in the reduced space. Despite their strengths, these methods often overlook the simultaneous consideration of all three crucial aspects—distance, conformity, and connectivity—limiting their effectiveness in segmenting overlapped sperm tails.

\section{Method}\label{sec:method}

Sperm image segmentation presents two principal challenges. The first challenge arises from the scarcity of labeled samples, often resulting in the undertraining of supervised learning models. The second challenge is the frequent occurrence of tail overlap and dye impurities, which complicates the independent segmentation of sperm and subsequently degrades the overall segmentation performance. 
In this section, we first examine how {\cname} effectively addresses the difficulty of overlapping tail segmentation. We then introduce the other components of our unsupervised method, {\name}, detailing how SAM is utilized for head segmentation, how dye interference is mitigated, and how the head and tail segments are subsequently spliced together.

\begin{figure}[t]
	\centering
	\begin{minipage}[h]{\linewidth}
		\centering
		\includegraphics[width=\linewidth]{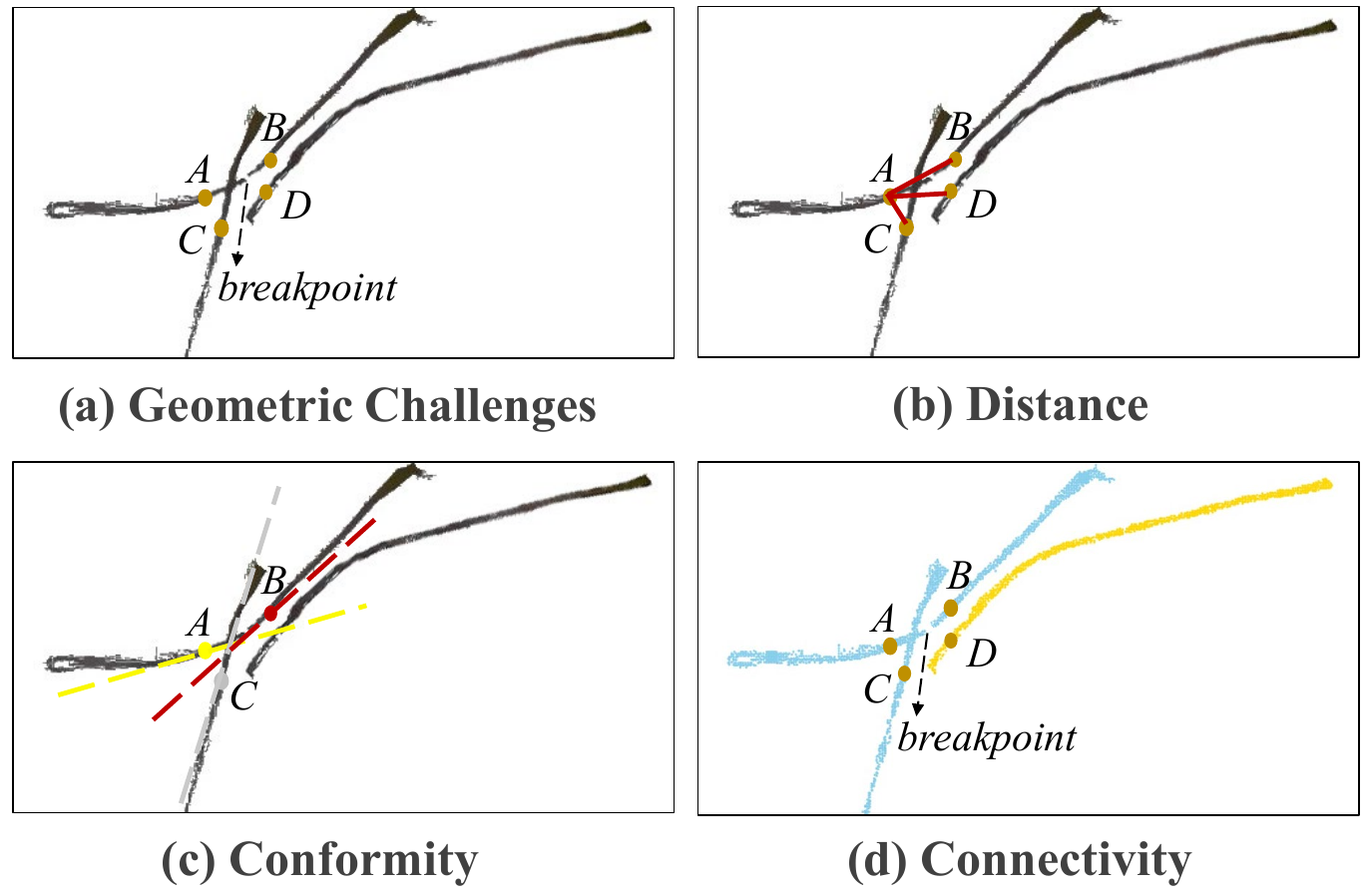}\\
	\end{minipage}
	\caption{(a) A pre-processed sperm tail image. The goal is to ensure that point A is similar to point B on the same tail, but not similar to points C and D on different tails. (b) Considering only distance, if B is similar to A, then C and D are also similar to A. (c) Considering conformity, point B is more similar to point A than point C based on the similarity of tangent angles. (d) Considering connectivity, among B, C, and D, only D is not in the same connected interval as A. Notably, small breakpoints can interfere with connectivity.}\label{fig:cluster}
	\vspace{-2mm}
\end{figure}

\begin{figure*}[ht]
	\centering
	\begin{minipage}[c]{\linewidth}
	\includegraphics[width=\linewidth]{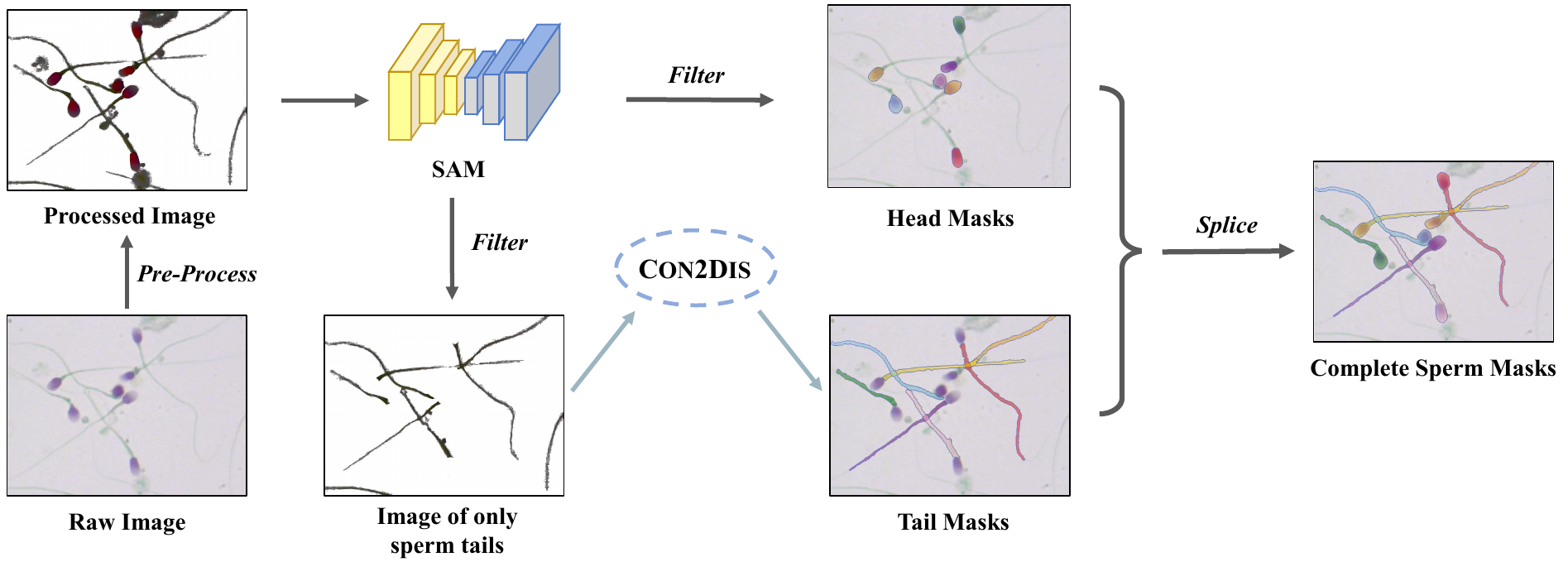}\\
    \end{minipage}
     	\vspace{-4mm}
	\caption{The pipeline of our {\name}. {\name} works in a ``decomposition-combination'' manner, utilizing SAM and {\cname} for segmenting heads and tails, respectively, and subsequently splicing them into complete masks.
    }\label{fig:pipeline}
    \vspace{-2mm}
\end{figure*}

    \subsection{Tail Segmentation with {\cname}}\label{sec:cluster}
    \textbf{Geometric Challenges in Clustering.} Viewing all sperm tail pixels as independent scattered points, clustering methods can be employed to group these points into different clusters, each representing an independent tail, thus achieving unsupervised instance segmentation of overlapping sperm tails. The clustering objective is to maximize intra-cluster similarity, ensuring that data points from the same sperm tail are grouped together, while minimizing inter-cluster similarity to segregate points from different tails. This section explores the complexities associated with achieving this objective. Referring to Figure~\ref{fig:cluster}(a), with point $A$ as the reference, the aim is to increase the similarity between points $A$ and $B$ while reducing the similarity between point $A$ and points $C$ and $D$. Some clustering methods, such as K-Means, rely solely on Euclidean distance as the measure of similarity between points, which can lead to problems. As illustrated in Figure~\ref{fig:cluster}(b), if point $B$ is considered similar to point $A$ based on distance, then points $C$ and $D$ would also be similar to point $A$. Manifold clustering methods, such as LSA~\cite{DBLP:conf/eccv/YanP06}, consider additional factors like conformity, such as the angle of the tangent line at the point. As shown in Figure~\ref{fig:cluster}(c), based on this consideration, point $B$ is closer to point $A$ than point $C$. However, for a sperm tail with variable width and uneven thickness, calculating tangent angles is challenging, particularly in overlapping regions. To exclude point $D$ from being similar to point $A$, connectivity must also be considered. As shown in Figure~\ref{fig:cluster}(d), different connected components are marked with different colors, allowing point $D$ to be excluded from $A$'s similar points. However, sperm images often contain breakpoints in the tail due to noise or impurities, which can interfere with connectivity-based judgments, leading to misclassification.
    
    In summary, a clustering method that simultaneously considers {\em distance}, {\em conformity}, and {\em connectivity} is required for effective segmentation of overlapping tails. The assessment of conformity and connectivity needs to be specially designed to handle the unique characteristics of sperm tails. \\

\noindent\textbf{Clustering Methodology.} For two sample points $x_i$ and $x_j$ in sperm tails, Our clustering method {\cname} defines {\em distance} $p_{ij}$ as:
\begin{equation}
p_{i j}= \begin{cases}1 & \text { if } x_i \in \operatorname{Knn}\left(x_j\right) \text { or } x_j \in \operatorname{Knn}\left(x_i\right) \\ 0 & \text {otherwise} \\ \end{cases}\;,\label{eq:pij}
\end{equation}
where $\operatorname{Knn}(x)$ denotes $K$ nearest neighbors of $x$. \\
The {\em connectivity} $r_{i j}$ is defined as:  
\begin{equation}
r_{i j}= \begin{cases}1 & \text { if } x_i, x_j \,  \text{are in the same connected component} \\ 0 & \text {otherwise} \\ \end{cases}\;.\label{eq:qij}
\end{equation}
To mitigate the interference caused by breakpoints, {\name} employs the DBSCAN~\cite{DBLP:conf/kdd/EsterKSX96} algorithm for connected component filtration, setting appropriate thresholds to prevent excessive segmentation while ensuring distinct tails are properly separated, as shown in Figure~\ref{fig:cluster}(d).

In general, the calculation of conformity relies on the construction of an undirected graph or affinity matrix. However, clusters near the intersection point can easily connect within the undirected graph, potentially corrupting the affinity matrix with poor pairwise affinity values, propagating misleading information among different clusters. Inspired by Spectral Multi-Manifold Clustering method~(SMMC)~\cite{DBLP:journals/tnn/WangJWZ11}, the {\em conformity} $q_{i j}$ is defined as:
\begin{equation}
q_{i j}=q\left(\Theta_i, \Theta_j\right)=\left(\prod_{l=1}^d \cos \left(\theta_l\right)\right)^o\;,\label{eq:rij}
\end{equation}
where $\Theta_i$ is the tangent space at $x_{i}$, and $o \in \mathrm{N}^{+}$ is an adjustable parameter. $0 \leq \theta_1 \leq, \cdots, \leq \theta_d \leq \pi / 2$ are a series of principal angles~\cite{golub2013matrix} between two tangent spaces $\Theta_i$ and $\Theta_j$, defined recursively as:
\begin{equation}
\cos \left(\theta_1\right)=\max _{\substack{u_1 \in \Theta_i, v_1 \in \Theta_j \\\left\|u_1\right\|=\left\|v_1\right\|=1}} u_1^T v_1\;,\label{eq:111}
\end{equation} and 
\begin{equation}
\cos \left(\theta_l\right)=\max _{\substack{u_l \in \Theta_i, v_l \in \Theta_j \\\left\|u_l\right\|=\left\|v_l\right\|=1}} u_l^T v_l, \quad l=2, \cdots, d \;.\label{eq:222}
\end{equation}
The tangent vectors $u_i$ and $v_i$, $i=1,\cdots, l$, are derived by training mixtures of probabilistic principal component analyzers using the EM~\cite{DBLP:journals/ijon/ArchambeauDV08} algorithm. Further details are provided in the supp. These multiple local tangent subspaces enhance the reliability of affinity values between pairwise points, particularly those located in overlapping areas of sperm tails.

The amalgamation of Eq.~\ref{eq:pij}, Eq.~\ref{eq:qij}, and Eq.~\ref{eq:rij} yields the similarity metric $w_{i j}$ defined as:
\begin{equation}
   w_{i j} = p_{i j} \cdot q_{i j} \cdot r_{i j} \;.\label{eq:wij}
\end{equation}
This metric satisfies the prerequisite clustering criteria, positing that a substantial $w_{i j}$ value is indicative of $x_i$ and $x_j$ being in close distance, possessing analogous tangent angles, and residing within the same connected component. With Eq.~\ref{eq:wij}, the affinity matrix $W$ is calculated as $W_{i j}= w_{i j}$ and the diagonal matrix $E$ is defined as $E_{i i}=\sum_j w_{i j}$. The clustering outcome is obtained by solving the first $k$ generalized eigenvectors $u_1, \cdots, u_k$ and the corresponding $k$ smallest eigenvalues of the generalized eigenproblem:
\begin{equation}
(E-W) u=\lambda E u \;.\label{eq:eigen}
\end{equation}
Finally, the K-means method is applied on the row vectors $U=\left[u_1, \cdots, u_k\right]$ to obtain the clusters.\\

\noindent\textbf{Skeletonization and Restoration of Tails.} 
The calculation of {\em conformity} $q_{i j}$ is complicated by variations in sperm tail thickness and noise interference. To address this, we apply skeletonization, reducing sperm tails to single-pixel-width lines, thereby improving the precision of angular assessments. Therefore, the clustering process produces a skeletal representation of the sperm tails, requiring a subsequent restoration phase to reconstruct the tails to their full form. This involves mapping each original tail point $x_i$ to its nearest skeletal counterpart $x_j$, facilitated by the distance similarity $p^{\prime}_{i j}$, calculated using Euclidean metrics as follows:
\begin{equation}
p^{\prime}_{i j} = \left\|x_i-x_j\right\|^2\;.\label{eq:pij2}
\end{equation}

A threshold $\gamma$ determines the allocation of point $x_i$ to the cluster associated with point $x_j$ if $p^{\prime}_{i j} < \gamma$. It is important to note that overlapping sperm tails may result in a single point being classified into multiple tails, accurately reflecting the biological complexity of sperm samples. Additionally,  skeletonization can simplify the computation of the distance $p_{i j}$ and the connectivity $r_{i j}$, while also facilitating the mask splicing of sperm heads and tails, as detailed in the following section. \\

\subsection{Sperm Segmentation with {\name}}\label{sec:SAM}
Following the segmentation of overlapping sperm tails using {\cname}, we propose {\name} for the comprehensive segmentation of complete sperm. As illustrated in Figure~\ref{fig:pipeline}, {\name} employs a ``decomposition-combination'' approach, comprising the following three main steps:
\begin{enumerate}
\item The Initial segmentation of pre-processed images using SAM to obtain masks for different sperm heads;
\item The segmentation of sperm tails with significant overlap using our clustering algorithm {\cname};
\item The combination of head and tail masks with a tailored mask splicing approach to produce complete masks.
\end{enumerate}

\noindent\textbf{Initial segmentation with SAM.} 
In sperm images, sperm heads generally appear as purple ovals, sperm tails as green tubular structures, and dye impurities as green irregular blocks. We utilize SAM for preliminary segmentation of sperm images, distinguishing these components based on color and shape. To enhance SAM's segmentation efficiency, we initiate a pre-processing phase for the input images, which includes adjustments to brightness, contrast, saturation, and resolution, along with edge sharpening and background whitening. SAM is then applied to these pre-processed images to generate various segmentation masks. 

We use shape and color indices, denoted as $SI$ and $CI$ respectively, to classify the mask types. For each mask $m$ segmented by SAM, we calculate its true area $S(m)$ and the area of its smallest circumscribed circle $S_{scc}(m)$. The shape index $SI$ is defined by their ratio:
\begin{equation}
    SI(m) = \frac{S(m)}{S_{scc}(m)}\;.\label{eq:SI}
\end{equation}
A larger $SI(m)$ indicates a clump-shaped mask, while a smaller $SI(m)$ suggests an elongated tubular shape. We set a threshold $\alpha$ ($0<\alpha<1$) for shape filtering. If $SI(m)<\alpha$, the mask $m$ is considered as an elongated tubular, indicative of a sperm tail; otherwise, it is deemed closer to a clump shape, suggestive of either a dye block or a sperm head. 

Additionally, we use $S_{pa}(m)$ to denote the ``purple area'' within the mask $m$, where ``purple area'' corresponds to the region in the raw image classified as purple by the RGB or HSV value. The color index $CI(m)$ is defined as:
\begin{equation}
    CI(m) = \frac{S_{pa}(m)}{S(m)}\;.\label{eq:CI}
\end{equation}

In raw sperm images, only sperm heads are dyed purple, while the rest is dyed green. We set another threshold $\beta$ ($0<\beta<1$) for color filtering, where a larger $\beta$ value indicates a greater presence of purple areas in the mask. The complete filtering method is defined as follows:
\begin{equation}
m= \begin{cases} 
\text{sperm head} & \text { if } SI(m)>\alpha \text { and } CI(m)>\beta \\ 
\text{dye block} & \text { if } SI(m)>\alpha \text { and } CI(m)<\beta \\ 
\text{sperm tail} & \text { if } SI(m)\leq \alpha \\ 
 \end{cases}\;.\label{eq:SICI}
\end{equation}

With Eq.~\ref{eq:SICI}, we effectively categorize all masks into three distinct types: heads, tails, and dye impurities. Following this classification, we isolate and store all masks identified as sperm heads separately. Subsequently, we remove the sperm heads and dye blocks from the \textit{raw} images, leaving \textit{only} the sperm tails as the foreground. Notably, sperm tails often overlap and intersect, a scenario where SAM typically struggles, often identifying these overlapping sections as a single entity rather than distinguishing them as separate structures. Given these limitations, we confine SAM's segmentation capabilities to sperm heads. \\

\noindent\textbf{Tail Segmentation with {\cname}.} We then apply {\cname} to further segment these images, which contain only overlapping tails, to obtain multiple independent masks for different sperm tails.    
This segmentation process may result in some sperm tails being fragmented into multiple sections due to occlusions by larger entities in raw images, such as heads or dye blocks, necessitating the subsequent detection and splicing of the masks corresponding to the same sperm tail. Fortunately, this is addressed in the head and tail splicing step. \\

\noindent\textbf{Splicing Masks of Heads and Tails.} In the previous steps, SAM is employed to segment sperm heads, followed by {\cname} for sperm tail segmentation. Finally, we implement a speicialized mask splicing approach to combine the corresponding head and tail masks associated with the same sperm. We fit each head mask into an ellipse and extract the line segment where the major axis of the ellipse lies, while the tail lines are obtained by skeletonization in {\cname}. The endpoints of these skeletonized tails and the major axes of the heads are identified, with particular emphasis on the terminal slopes at these endpoints. Our approach is based on the hypothesis that a minimal distance between endpoints and a similarity in terminal slopes suggest a higher likelihood of the endpoints belonging to the same sperm entity. For any two distinct endpoints $x_i$ and $x_j$ that do not originate from the same line, we first evaluate their distance similarity $p^{\prime}_{i j}$ using Eq.~\ref{eq:pij2}. We then assess their angular similarity $q^{\prime}_{i j}$, derived from the tangent slopes at these endpoints, as follows:
\begin{equation}
q^{\prime}_{i j} = \lvert  s\left( x_i \right) - s\left( x_j \right)\rvert\;,\label{eq:qij2}
\end{equation}
where $s(\cdot)$ denotes the slope value, and 
$\lvert\cdot\rvert$ denotes the absolute value. By instituting distance threshold $\lambda_{1}$ and angle threshold $\lambda_{2}$, we proceed to connect $x_i$ and $x_j$ with a linear path, if $p^{\prime}_{i j}< \lambda_{1}$ and $q^{\prime}_{i j}< \lambda_{2}$ simultaneously. If multiple endpoints meet this criterion for a single anchor endpoint, the endpoint closest in \textit{angle} is preferred. In addition to head-tail splicing, our mask splicing approach also matches endpoints from different tails, as sperm tails may be broken into several segments due to various disturbances. Successful matching results in the assembly of complete sperm masks, effectively integrating the head and tail components.

\section{Experiment}\label{sec:exp}

We evaluate the effectiveness of our proposed method, {\name}, on a sperm dataset obtained from clinical practice. The visual results demonstrate that {\name} provides superior segmentation of overlapping sperm tails compared to existing methods. Additional experimental settings and results are provided in the supplementary materials.

\subsection{Experimental Setups}
\begin{figure*}[t]
	\centering
	\begin{minipage}[c]{\linewidth}
	\includegraphics[width=\linewidth]{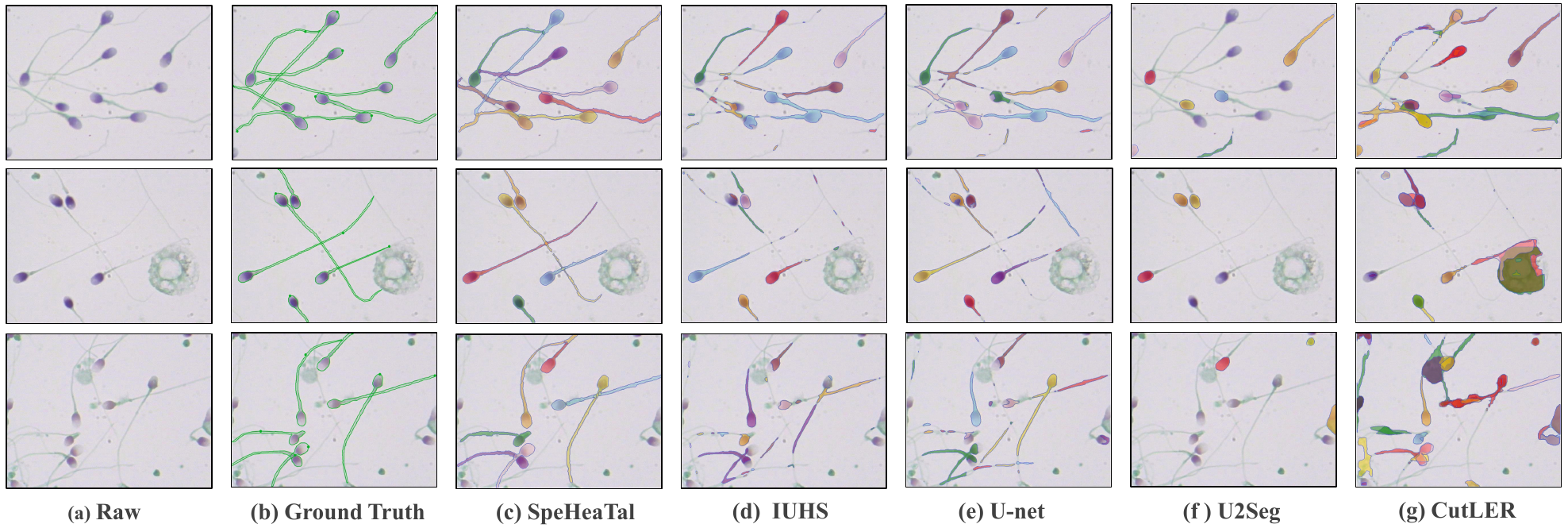}\\
    \end{minipage}
    \vspace{-4mm}
	\caption{Visualization of segmentation results obtained by different methods. More results can be found in the supp.
    }\label{fig:visual}
    \vspace{-2mm}
\end{figure*}
\noindent\textbf{Datasets.} 
The sperm datasets used in previous works~\cite{DBLP:journals/npl/LvYQLZZ22,DBLP:conf/cbms/MelendezCR21,Iqbal2020Deep} predominantly feature images with independent sperm, which do not adequately represent the cross-overlapping phenomena commonly encountered in clinical practice. To address this gap, we collaborated with several leading hospitals to obtain approximately 2,000 unlabeled sperm images from clinical settings for model parameter adjustment. Additionally, 240 sperm images were annotated by andrology experts, with each annotation verified and approved by at least three senior experts to ensure accuracy for model performance testing. The original images are all 720$\times$540 pixels in size. For those models requiring labeled samples for training, we divided the 240 labeled images into a 3:1 ratio, with 180 images used for their training and the remaining 60 images reserved for performance evaluation. \\

\noindent\textbf{Evaluation Metrics.} Following the methodologies outlined in~\cite{DBLP:journals/corr/abs-2202-05273,Costa2019Evaluating,baccouche2021connected,shamir2019continuous}, we employ mean Intersection over Union~(mIoU) and mean Dice index~(mDice), two standard metrics in instance segmentation, to evaluate model performance. Both mIoU and mDice assess performance by measuring the degree of overlap between the ground truth areas ($A_{i}, i=1,\cdots,N$) and the predicted areas ($B_{i}, i=1,\cdots,N$). A higher degree of overlap corresponds to a larger index value and better model performance. The specific definitions are as follows:
\begin{equation}
\text{mIoU} = \frac{1}{N} \sum_{i=1}^{N} \left( \frac{|A_i \cap B_i|}{|A_i \cup B_i|} \right),\label{eq:mIoU}
\end{equation}
\begin{equation}
\text{mDice} = \frac{1}{N} \sum_{i=1}^{N} \left( \frac{2 \times |A_i \cap B_i|}{|A_i| + |B_i|} \right).\label{eq:mDice}
\end{equation}

Prior to calculating mIoU and mDice, we establish the optimal pairing relationship by comparing the IoU ratios between each instance in Ground Truth and the predicted results. Notably, the segmentation outcomes of Ground Truth and certain methods may not correspond on a one-to-one basis, potentially leading to a disparity in quantities. In such cases, we apply a filtering criterion based on selecting pairs with an optimal pairing IoU ratio for calculation.\\

\noindent\textbf{Comparison Methods.}
We categorize our comparison methods into two groups. The first group consists of supervised learning methods based on frameworks such as U-Net~\cite{DBLP:conf/miccai/RonnebergerFB15}. Previous sperm segmentation methods fall into this category, including IUHS~\cite{DBLP:journals/npl/LvYQLZZ22} and CN2UA~\cite{DBLP:conf/cbms/MelendezCR21}. Since these methods do not provide source codes, we reproduced them based on the original papers. The second group comprises unsupervised instance segmentation methods, including SAM~\cite{Alexander2023SAM}, U2Seg~\cite{DBLP:journals/corr/abs-2312-17243}, and CutLER~\cite{DBLP:conf/cvpr/0007GYM23}, for which we directly used the authors' provided codes. \\

\noindent\textbf{Implementation details.}
Our method utilizes the SAM model in ``everything mode'' to segment microscopic sperm images. The images are standardized to a resolution of 720$\times$540 pixels to ensure consistency across the dataset. According to the parameter settings described in the methodology section, the shape filtering threshold $\alpha = 0.25$, color filtering threshold $\beta = 0.4$, and point allocation threshold $\gamma = 5$ pixels. The distance threshold $\lambda_{1} = 30$ pixels and angle threshold $\lambda_{2} = 35^\circ$ are determined to effectively splice different parts of each sperm. Codes are available at \textit{https://www.github.com/shiy19/SpeHeaTal}.

\subsection{Performance Evaluation}

\begin{table}[t]
	\centering
	\tabcolsep 8pt
	\begin{tabular}{c|cc}
		\addlinespace
		\toprule

		Criteria & mIOU & mDice \\
		\toprule
  		U-net~\cite{DBLP:conf/miccai/RonnebergerFB15}	& 49.21 & 62.17 \\
    		IUHS~\cite{DBLP:journals/npl/LvYQLZZ22} & 47.56 & 58.46  \\ 
      		CN2UA~\cite{DBLP:conf/cbms/MelendezCR21} & 19.26 & 32.94  \\ 
          	 SAM~\cite{Alexander2023SAM}  & 34.17 & 46.75  \\
		U2Seg~\cite{DBLP:journals/corr/abs-2312-17243}  & 16.44 & 28.86  \\
		CutLER~\cite{DBLP:conf/cvpr/0007GYM23}	& 31.21 & 44.17 \\
		\midrule
	    {\name} & {\bf 72.21} & {\bf 81.07} \\
		\bottomrule
	\end{tabular}
    	\caption{mIOU and mDice of different segmentation methods on our sperm dataset. The best results are in bold.}\label{table:exp}
\vspace{-2mm}
\end{table}

The experimental results on our sperm dataset are presented in Table~\ref{table:exp}. All comparison methods and our proposed {\name} are tested on pre-processed images. Supervised methods, such as U-Net, IUHS, and CN2UA, require a large amount of labeled data for training and exhibit limited effectiveness on sperm datasets lacking large-scale labeled data. Consequently, these methods often simplify the sperm segmentation task, rendering them ineffective on our real clinical sperm dataset. In contrast, unsupervised instance segmentation methods, including SAM, U2Seg, and CutLER, often rely on texture and structural patterns in images. However, sperm imagery significantly deviates from typical natural images, characterized by high contrast, minimal texture, and disruptions due to dye blocks. These unique attributes complicate the learning process for these models. Although some methods utilize robust data augmentation to induce variability, such enhancements risk obliterating critical features in sperm imagery, such as tail position and morphology, thus impairing model performance. Furthermore, the frequent presence of numerous elongated tails and their irregular distribution across the image exacerbates the difficulty of characterization. All comparison methods struggle with processing these overlapping structures. 

In contrast, our method {\name} exhibits a remarkable ability to accurately segment sperm tails, even in scenarios with significant overlap and dye impurities, and effectively identifies masks for multiple independent sperm. This capability results in superior performance across the evaluated metrics. Visualizations of segmentation results obtained by different methods are shown in Figure~\ref{fig:visual}. 

\subsection{Ablation Study}
\begin{table}[t]
	\centering

	\tabcolsep 1pt
	\begin{tabular}{c|ccc|cc}
		\addlinespace
		\toprule

		Method & Distance & Conformity & Connectivity & mIOU & mDice \\
		\toprule
		K-Means & \ding{51}& \ding{55}& \ding{55} & 14.63 & 26.94  \\
		AHC	& \ding{51}& \ding{55}& \ding{51} & 27.76 & 38.17  \\ 
		LSA & \ding{51}& \ding{51}& \ding{55} & 37.61 & 45.92 \\
  	 SC & \ding{51}& \ding{51}& \ding{55} & 46.86 & 56.33  \\ 
		\midrule
	    {\cname} & \ding{51}& \ding{51}& \ding{51} & {\bf 72.21} & {\bf 81.07} \\
		\bottomrule
	\end{tabular}
 	\caption{mIOU and mDice of different clustering methods on our sperm dataset. The best results are in bold.}\label{table:3}
\vspace{-2mm}
\end{table}

We further investigate the impact of using alternative clustering techniques, such as K-Means,  Agglomerative Hierarchical Clustering~(AHC), Local Subspace Affinity~(LSA), and Spectral Clustering~(SC), in place of {\cname}. The experimental results are detailed in Table~\ref{table:3}.
All methods are applied solely to sperm tail segmentation, with SAM used for sperm head segmentation. Sperm tail imagery presents intricate geometric challenges, necessitating a multi-faceted approach to clustering for precise segmentation of overlapping sperms. This approach encompasses three critical perspectives: distance, conformity, and connectivity. The comparison methods inadequately address all these factors. While some methods consider certain aspects, such as SC considering conformity, they still fall short. For instance, the cross-overlap at the tail significantly disrupts the calculation of the affinity matrix in SC. Consequently, our tailored {\cname} achieves the best performance.

\begin{figure}[t]
	\centering
	\begin{minipage}[c]{\linewidth}
	\includegraphics[width=\linewidth]{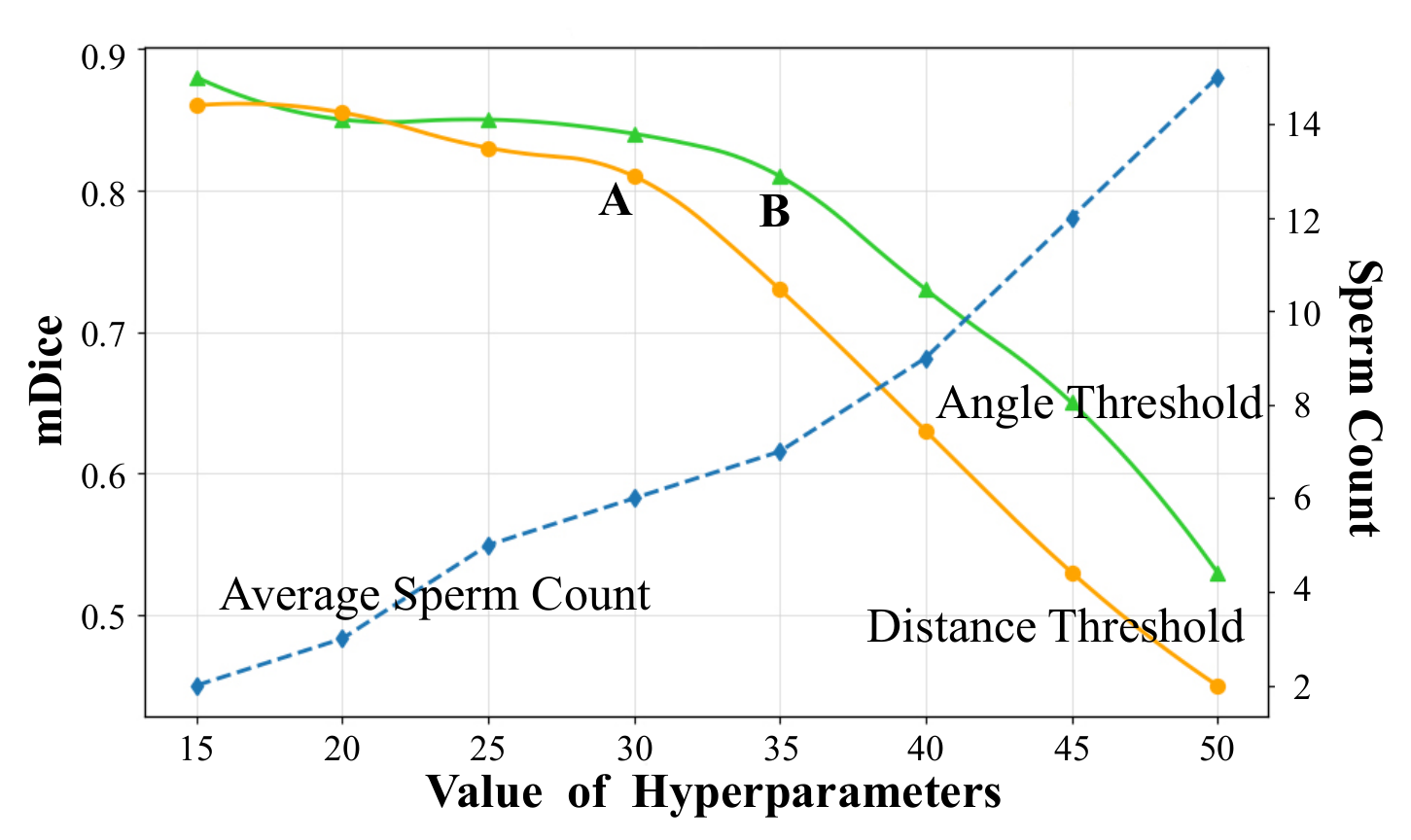}\\
    \end{minipage}
        \vspace{-4mm}
	\caption{The impact of distance threshold $\lambda_{1}$ and angle threshold $\lambda_{2}$ on {\name}.
    }\label{fig:abl}
    \vspace{-2mm}
\end{figure}
We also conduct an additional ablation study to explore the impact of two hyperparameters, the distance threshold $\lambda_{1}$ and the angle threshold $\lambda_{2}$, in our mask splicing approach. The results are shown in Figure~\ref{fig:abl}. We also visualize the average \textit{sperm count} detected in each image for different \textit{distance thresholds}. As $\lambda_{1}$ and $\lambda_{2}$ increase, the model relaxes the screening conditions, leading to more sperm with head-to-tail mismatches and incomplete head-to-tail matches being counted, resulting in a continued decline in mDice. Additionally, due to the looser restrictions, the number of successfully detected sperm in each image gradually increases. Based on the graph, we select points $A$ and $B$, where the ``elbow'' inflection point of the mDice parameter curve appears, as appropriate hyperparameter values. These hyperparameter values represent an optimal balance between detection accuracy and detection completeness.

\section{Conclusion}\label{sec:con}
The challenges of tail overlap and dye impurities pose significant obstacles to the unsupervised segmentation of sperm images. In this work, we first conduct a geometric analysis, revealing that the effective segmentation of overlapping tails requires the simultaneous consideration of Conformity, Connectivity, and Distance. Leveraging these insights, we develop a clustering algorithm, {\cname}, specifically tailored to address the segmentation of overlapping tails. Building on this foundation, we introduce {\name}, a comprehensive sperm segmentation method that employs a ``decomposition-combination'' approach. {\name} utilizes SAM for head segmentation and impurity removal, while {\cname} is used for tail segmentation. These segments are then integrated into complete masks through a tailored mask splicing technique. Experimental results on an in-house clinical sperm dataset demonstrate that {\name} outperforms existing methods, particularly in images with significant overlap. \\


\noindent \textbf{Limitations.} {\name} encounters difficulties when processing images with extremely complex tail overlaps, particularly in scenarios where more than ten sperm are entangled—situations that even human experts struggle to resolve. Therefore, adequate dilution of semen samples remains necessary in clinical practice to ensure manageable analysis conditions.

\section{Acknowledgments}
This work is supported by the NSFC Project~(No. 62176117, 62476123), the National Natural Science Foundation of China~(U22A20277), and Jiangsu Provincial Medical Key Discipline Cultivation Unit~(JSDW202215).

\bibliography{aaai25}

\end{document}